# Spread–Spectrum Based on Finite Field Fourier Transforms

H.M. de Oliveira, J.P.C.L. Miranda, R.M. Campello de Souza

CODEC – Grupo de Pesquisas em Comunicações
Departamento de Eletrônica e Sistemas – CTG – UFPE
C.P. 7800, 50711-970, Recife – PE, Brazil
e-mail: hmo@npd.ufpe,br, jump@elogica.com.br, ricardo@npd.ufpe.br

***ABSTRACT***

Spread-spectrum systems are presented, which are based on Finite Field Fourier Transforms. Orthogonal spreading sequences defined over a finite field are derived. New digital multiplex schemes based on such spread-spectrum systems are also introduced, which are multilevel Coding Division Multiplex. These schemes termed Galois-field Division Multiplex (GDM) offer compact bandwidth requirements because only leaders of cyclotomic cosets are needed to be transmitted.

## 1. Introduction

CDMA is becoming the most popular multiple access scheme. It is well known that spread spectrum waveforms provide multiple-access capability and low probability of interception [QUAL 92], [MAS 95]. In this paper a spread-spectrum system is introduced which is based on the Finite Field Fourier Transform (FFFT) [POL 71]. The FFFT has successfully been applied to perform discrete convolution, image processing, coding and cryptography among many applications [REE et al. 77], [REE & TRU 79], [BLA 79].

## 2. Galois-Fourier Spreading Sequences

Let $v = (v_0, v_1, ..., v_{N-1})$, $v_i \in GF(p)$, be a "user"- vector of length N and let $V = (V_0, V_1, ..., V_{N-1})$, $V_k \in GF(p^m)$, be the Galois spectrum of v, i.e.

$$V_K = \sum_{i=0}^{N-1} v_i \alpha^{ik} \qquad (1)$$

where $\alpha$ is an element of multiplicative order N over the extension field $GF(p^m)$.

Example 1: Consider a primitive polynomial over GF(2), say $p(x) = x^4 + x + 1$. Let $\alpha$ be a root of p(x), i.e., $p(\alpha) = 0$. Then $\alpha$ is a primitive element of GF(16) and $N = ord(\alpha) = 2^4 - 1 = 15$. The extension field GF(16) showed in table I can be generated as usual [McE 87].

Table I – Galois Field GF(16).

| i | $\alpha^i$ | associated vector | $ord(\alpha^i)$ | minimal polynomial |
|---|---|---|---|---|
| 0 | $\alpha^0 = 1$ | (0,0,0,1) | 1 | $x + 1$ |
| 1 | $\alpha^1 = \alpha$ | (0,0,1,0) | 15 | $x^4 + x + 1$ |
| 2 | $\alpha^2 = \alpha^2$ | (0,1,0,0) | 15 | $x^4 + x + 1$ |
| 3 | $\alpha^3 = \alpha^3$ | (1,0,0,0) | 5 | $x^4 + x^3 + x^2 + x + 1$ |
| 4 | $\alpha^4 = \alpha + 1$ | (0,0,1,1) | 15 | $x^4 + x + 1$ |
| 5 | $\alpha^5 = \alpha^2 + \alpha$ | (0,1,1,0) | 3 | $x^2 + x + 1$ |
| 6 | $\alpha^6 = \alpha^3 + \alpha^2$ | (1,1,0,0) | 5 | $x^4 + x^3 + x^2 + x + 1$ |
| 7 | $\alpha^7 = \alpha^3 + \alpha + 1$ | (1,0,1,1) | 15 | $x^4 + x^3 + 1$ |
| 8 | $\alpha^8 = \alpha^2 + 1$ | (0,1,0,1) | 15 | $x^4 + x + 1$ |
| 9 | $\alpha^9 = \alpha^3 + \alpha$ | (1,0,1,0) | 5 | $x^4 + x^3 + x^2 + x + 1$ |
| 10 | $\alpha^{10} = \alpha^2 + \alpha + 1$ | (0,1,1,1) | 3 | $x^2 + x + 1$ |
| 11 | $\alpha^{11} = \alpha^3 + \alpha^2 + \alpha$ | (1,1,1,0) | 15 | $x^4 + x^3 + 1$ |
| 12 | $\alpha^{12} = \alpha^3 + \alpha^2 + \alpha + 1$ | (1,1,1,1) | 5 | $x^4 + x^3 + x^2 + x + 1$ |
| 13 | $\alpha^{13} = \alpha^3 + \alpha^2 + 1$ | (1,1,0,1) | 15 | $x^4 + x^3 + 1$ |
| 14 | $\alpha^{14} = \alpha^3 + 1$ | (1,0,0,1) | 15 | $x^4 + x^3 + 1$ |

Once $N \mid p^m - 1$, one could choose any non-primitive element with order equals to 1, 3 or 5. Let $\alpha$ be a primitive element of GF(16), thus $ord(\alpha) = 15$. Now let us define the spreading sequences, from now on denoted Galois-Fourier carriers over GF(16) according to $\{\alpha^{ik}\}_{k=0}^{N-1}$:

$\alpha^{0k} = (1, 1, 1, 1, 1, 1, 1, 1, 1, 1, 1, 1, 1, 1, 1)$
$\alpha^{1k} = (1, \alpha, \alpha^2, \alpha^3, \alpha^4, \alpha^5, \alpha^6, \alpha^7, \alpha^8, \alpha^9, \alpha^{10}, \alpha^{11}, \alpha^{12}, \alpha^{13}, \alpha^{14})$
$\alpha^{2k} = (1, \alpha^2, \alpha^4, \alpha^6, \alpha^8, \alpha^{10}, \alpha^{12}, \alpha^{14}, \alpha, \alpha^3, \alpha^5, \alpha^7, \alpha^9, \alpha^{11}, \alpha^{13})$
$\alpha^{3k} = (1, \alpha^3, \alpha^6, \alpha^9, \alpha^{12}, 1, \alpha^3, \alpha^6, \alpha^9, \alpha^{12}, 1, \alpha^3, \alpha^6, \alpha^9, \alpha^{12})$
$\alpha^{4k} = (1, \alpha^4, \alpha^8, \alpha^{12}, \alpha, \alpha^5, \alpha^9, \alpha^{13}, \alpha^2, \alpha^6, \alpha^{10}, \alpha^{14}, \alpha^3, \alpha^7, \alpha^{11})$
$\alpha^{5k} = (1, \alpha^5, \alpha^{10}, 1, \alpha^5, \alpha^{10}, 1, \alpha^5, \alpha^{10}, 1, \alpha^5, \alpha^{10}, 1, \alpha^5, \alpha^{10})$
$\alpha^{6k} = (1, \alpha^6, \alpha^{12}, \alpha^3, \alpha^9, 1, \alpha^6, \alpha^{12}, \alpha^3, \alpha^9, 1, \alpha^6, \alpha^{12}, \alpha^3, \alpha^9)$
$\alpha^{7k} = (1, \alpha^7, \alpha^{14}, \alpha^6, \alpha^{13}, \alpha^5, \alpha^{12}, \alpha^4, \alpha^{11}, \alpha^3, \alpha^{10}, \alpha^2, \alpha^9, \alpha, \alpha^8)$
$\alpha^{8k} = (1, \alpha^8, \alpha, \alpha^9, \alpha^2, \alpha^{10}, \alpha^3, \alpha^{11}, \alpha^4, \alpha^{12}, \alpha^5, \alpha^{13}, \alpha^6, \alpha^{14}, \alpha^7)$
$\alpha^{9k} = (1, \alpha^9, \alpha^3, \alpha^{12}, \alpha^6, 1, \alpha^9, \alpha^3, \alpha^{12}, \alpha^6, 1, \alpha^9, \alpha^3, \alpha^{12}, \alpha^6)$
$\alpha^{10k} = (1, \alpha^{10}, \alpha^5, 1, \alpha^{10}, \alpha^5, 1, \alpha^{10}, \alpha^5, 1, \alpha^{10}, \alpha^5, 1, \alpha^{10}, \alpha^5)$
$\alpha^{11k} = (1, \alpha^{11}, \alpha^7, \alpha^3, \alpha^{14}, \alpha^{10}, \alpha^6, \alpha^2, \alpha^{13}, \alpha^9, \alpha^5, \alpha, \alpha^{12}, \alpha^8, \alpha^4)$
$\alpha^{12k} = (1, \alpha^{12}, \alpha^9, \alpha^6, \alpha^3, 1, \alpha^{12}, \alpha^9, \alpha^6, \alpha^3, 1, \alpha^{12}, \alpha^9, \alpha^6, \alpha^3)$
$\alpha^{13k} = (1, \alpha^{13}, \alpha^{11}, \alpha^9, \alpha^7, \alpha^5, \alpha^3, \alpha, \alpha^{14}, \alpha^{12}, \alpha^{10}, \alpha^8, \alpha^6, \alpha^4, \alpha^2)$
$\alpha^{14k} = (1, \alpha^{14}, \alpha^{13}, \alpha^{12}, \alpha^{11}, \alpha^{10}, \alpha^9, \alpha^8, \alpha^7, \alpha^6, \alpha^5, \alpha^4, \alpha^3, \alpha^2, \alpha)$.

Each carrier should be interpreted as a spreading sequence. These spreading sequences can be viewed as some kind of generalisation of classical synchronous orthogonal spreading-spectrum sequences. Let us now investigate the correlation properties of Galois-Fourier carriers.

Propositon 1: The correlation property for Galois-Fourier carriers is given by:

$$R(i-t)=\sum_{k=0}^{N-1}\alpha^{ik}\alpha^{-tk}=\begin{cases}N, i\equiv t \pmod N\\ 0,\ \text{other cases.}\end{cases}$$

Proof: The autocorrelation function (ACF) is

$$R(i,t)=\sum_{k=0}^{N-1}\alpha^{ik}\alpha^{-tk}=\sum_{k=0}^{N-1}\alpha^{(i-t)k}=R(i-t)$$

Supposing now that i ≡ t (mod N), the maximum value of the ACF of sequences can be derived:

$$R(0)=\sum_{k=0}^{N-1}\alpha^{ik}\alpha^{-tk}=N$$

Thus, the ACF can be improved by increasing the extension field since N = ord(α) | $p^m$ – 1. Therefore, better autocorrelation is obtained as the number of users increase, which is a nice property. On the other hand, if i – t = j ≠ 0 (mod N), then the cross-correlation between Galois-Fourier carriers is therefore

$$R(j)=\sum_{k=0}^{N-1}\alpha^{jk}=\frac{1-\alpha^{jN}}{1-\alpha^{j}}=0,$$

since ord(α) = N. ■

A N-user mux has one carrier per channel. The requirements to achieve Welch's lower bound according [MAS & MIT 91] are provided by these sequences. The matrix [{α}] presents both orthogonal rows and orthogonal columns having the same "energy". In the absence of noise, there is no cross-talk from any user to any other one.

# 3. Multiplexing by Transform Spread-Spectrum

The Galois-Fourier carriers derived in example 1 can be used to implement a 15-user binary multiplex over the extension field GF(16). Supposing that the "user"-vector is

$$v = (v_0, v_1, ..., v_{14}) = (0, 1, 1, 0, 1, 0, 0, 0, 1, 0, 1, 1, 0, 0, 1),$$

where each component $v_i$ corresponds to a binary symbol of duration T for the $i^{th}$-user. Computing the FFFT using (1) yields:

$$V = (V_0, V_1, ..., V_{14}) = (1, 0, 0, \alpha^{10}, 0, \alpha^5, \alpha^5, \alpha^{10}, 0, \alpha^5, \alpha^{10}, \alpha^5, \alpha^{10}, \alpha^{10}, \alpha^5).$$

The spread-signal is generated by a componentwise multiplication between the signal v, $v_i$ ∈ GF(2) and the Galois-Fourier carriers defined over GF(16). The transform vector can be derived from the scheme showed in figure 1.

The componentwise product of an information symbol from the user by its corresponding Galois-Fourier carrier furnishes the spread-vector. The output is exactly the Finite Field Fourier Transform of the "user"-vector v, so it contains all the information about all channels [deO et al. 98]. Each coefficient $V_k$ of the Galois spectrum has duration T / N.

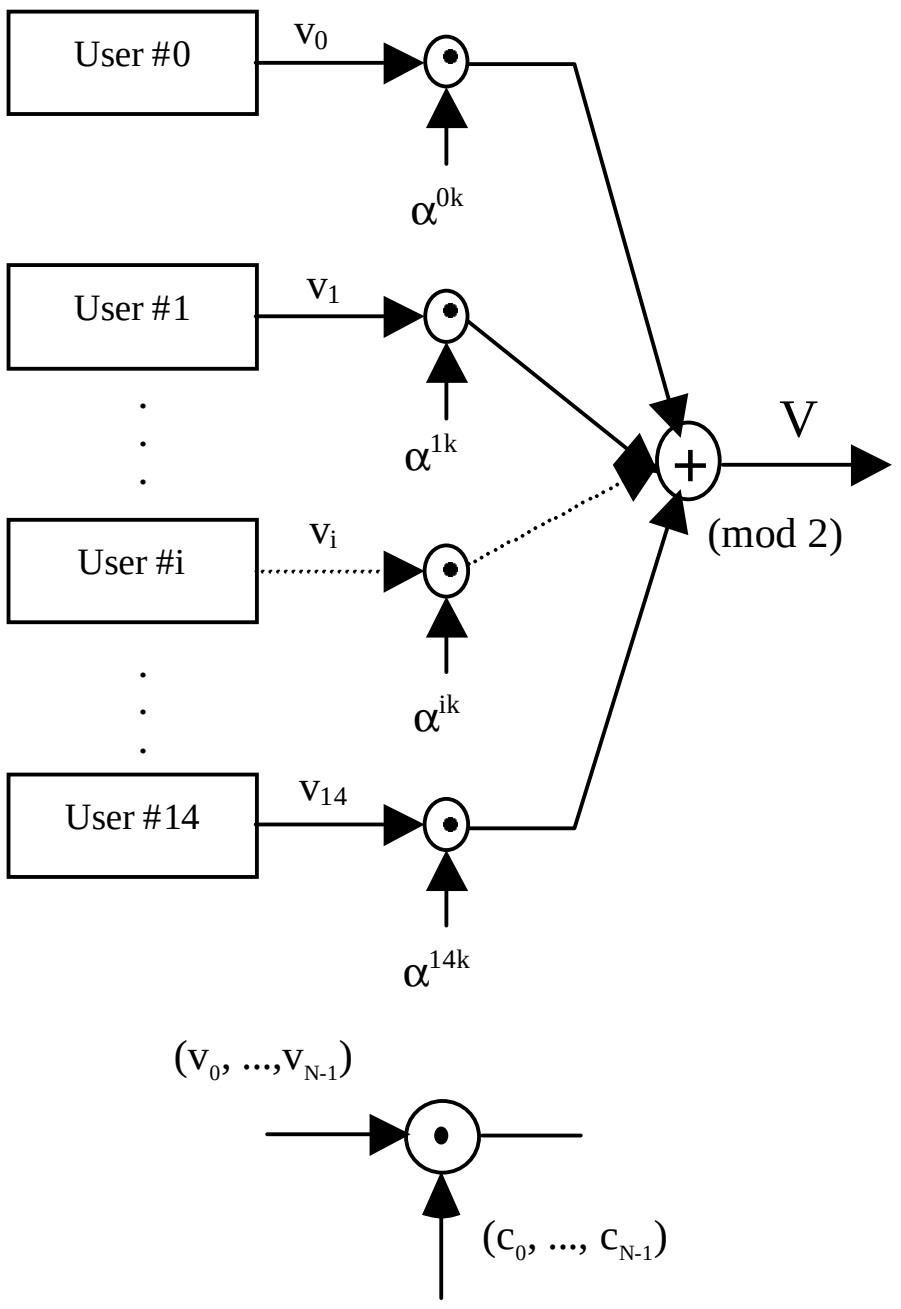

Fig. 1 – Galois-Fourier spreading sequences over GF(16).

Discrete Fourier Transforms (DFT) have long been applied in multicarrier modulation schemes [e.g. URI & CAR 99] referred to as Orthogonal Frequency Multiplexing (OFDM). Despite some similarities between OFDM and GDM, the true nature of these schemes is quite different. The first one is analogue, discrete-time, frequency-division while the later performs a digital Galois division.

Table II shows the spread signal vectors for each user. The addition of elements in the same line yields the Galois spectrum V of the input (column) binary signal.

Table II – Spread-vector per user.

| user # | | | | | | | | | | | | | | | V |
|---|---|---|---|---|---|---|---|---|---|---|---|---|---|---|---|
| 0 | 1 | 2 | 3 | 4 | 5 | 6 | 7 | 8 | 9 | 10 | 11 | 12 | 13 | 14 | |
| 0 | 1 | 1 | 0 | 1 | 0 | 0 | 0 | 1 | 0 | 1 | 1 | 0 | 0 | 1 | $V_0$ |
| 0 | $\alpha$ | $\alpha^2$ | 0 | $\alpha^4$ | 0 | 0 | 0 | $\alpha^8$ | 0 | $\alpha^{10}$ | $\alpha^{11}$ | 0 | 0 | $\alpha^{14}$ | $V_1$ |
| 0 | $\alpha^2$ | $\alpha^4$ | 0 | $\alpha^8$ | 0 | 0 | 0 | $\alpha$ | 0 | $\alpha^5$ | $\alpha^7$ | 0 | 0 | $\alpha^{13}$ | $V_2$ |
| 0 | $\alpha^3$ | $\alpha^6$ | 0 | $\alpha^{12}$ | 0 | 0 | 0 | $\alpha^9$ | 0 | 1 | $\alpha^3$ | 0 | 0 | $\alpha^{12}$ | $V_3$ |
| 0 | $\alpha^4$ | $\alpha^8$ | 0 | $\alpha$ | 0 | 0 | 0 | $\alpha^2$ | 0 | $\alpha^{10}$ | $\alpha^{14}$ | 0 | 0 | $\alpha^{11}$ | $V_4$ |
| 0 | $\alpha^5$ | $\alpha^{10}$ | 0 | $\alpha^5$ | 0 | 0 | 0 | $\alpha^{10}$ | 0 | $\alpha^5$ | $\alpha^{10}$ | 0 | 0 | $\alpha^{10}$ | $V_5$ |
| 0 | $\alpha^6$ | $\alpha^{12}$ | 0 | $\alpha^9$ | 0 | 0 | 0 | $\alpha^3$ | 0 | 1 | $\alpha^6$ | 0 | 0 | $\alpha^9$ | $V_6$ |
| 0 | $\alpha^7$ | $\alpha^{14}$ | 0 | $\alpha^{13}$ | 0 | 0 | 0 | $\alpha^{11}$ | 0 | $\alpha^{10}$ | $\alpha^2$ | 0 | 0 | $\alpha^8$ | $V_7$ |
| 0 | $\alpha^8$ | $\alpha$ | 0 | $\alpha^2$ | 0 | 0 | 0 | $\alpha^4$ | 0 | $\alpha^5$ | $\alpha^{13}$ | 0 | 0 | $\alpha^7$ | $V_8$ |
| 0 | $\alpha^9$ | $\alpha^3$ | 0 | $\alpha^6$ | 0 | 0 | 0 | $\alpha^{12}$ | 0 | 1 | $\alpha^9$ | 0 | 0 | $\alpha^6$ | $V_9$ |
| 0 | $\alpha^{10}$ | $\alpha^5$ | 0 | $\alpha^{10}$ | 0 | 0 | 0 | $\alpha^5$ | 0 | $\alpha^{10}$ | $\alpha^5$ | 0 | 0 | $\alpha^5$ | $V_{10}$ |
| 0 | $\alpha^{11}$ | $\alpha^7$ | 0 | $\alpha^{14}$ | 0 | 0 | 0 | $\alpha^{13}$ | 0 | $\alpha^5$ | $\alpha$ | 0 | 0 | $\alpha^4$ | $V_{11}$ |
| 0 | $\alpha^{12}$ | $\alpha^9$ | 0 | $\alpha^3$ | 0 | 0 | 0 | $\alpha^6$ | 0 | 1 | $\alpha^{12}$ | 0 | 0 | $\alpha^3$ | $V_{12}$ |
| 0 | $\alpha^{13}$ | $\alpha^{11}$ | 0 | $\alpha^7$ | 0 | 0 | 0 | $\alpha^{14}$ | 0 | $\alpha^{10}$ | $\alpha^8$ | 0 | 0 | $\alpha^2$ | $V_{13}$ |
| 0 | $\alpha^{14}$ | $\alpha^{13}$ | 0 | $\alpha^{11}$ | 0 | 0 | 0 | $\alpha^7$ | 0 | $\alpha^5$ | $\alpha^4$ | 0 | 0 | $\alpha$ | $V_{14}$ |

## *3.1 Selecting Cyclotomic Leaders*

According to Mœbius' formula the number of distinct irreducible polynomials of degree k over GF(p) is given by

$$I_k(p) = \frac{1}{k}\sum_{d|k}\mu(d).p^{k/d},$$

where $\mu$ is the Mœbius' function [McE 87]. The number of cyclotomic classes of the FFFT over GF(16) is therefore:

$$v_F = \sum_{k|m} I_k(p) - 1 = 5$$

Factoring $x^{15} - 1$ over GF(16), cyclotomic cosets and associated minimal polynomials can be found in table III.

Table III – Cyclotomic cosets and associated minimal polynomials.

| cyclotomic cosets | minimal polynomial |
|---|---|
| C(0) = (0) | $x + 1$ |
| C(1) = (1, 2, 4, 8) | $x^4 + x + 1$ |
| C(3) = (3, 6, 12, 9) | $x^4 + x^3 + x^2 + x + 1$ |
| C(5) = (5, 10) | $x^2 + x + 1$ |
| C(7) = (7, 14, 13, 11) | $x^4 + x^3 + 1$ |

Data compression is achieved by transmitting only the leader of each cyclotomic coset. Thus:

$$V_{comp} = (V_0, V_1, V_3, V_5, V_7) = (1, 0, \alpha^{10}, \alpha^5, \alpha^{10}).$$

Definition 1: The GDM bandwidth compactness factor is defined as $\gamma_{cc} = N / \nu$, where N is the number of users and $\nu$ is the number of cyclotomic cosets associated with a FFFT spectrum. ■

For the case under study, $\gamma_{cc} = 15 / 5 = 3$. If the full-vector V is transmitted, and assuming $B_1$ Hz the bandwidth required for individual user, GDM bandwidth requirements will be $B_{GDM} = NB_1$ Hz, the same result as TDM. Conversely, when compression is carried out, the bandwidth needed is $B_{GDM\ Comp} = NB_1 / \gamma_{cc} = B_{TDM} / 3$.

## *3.2 Combining MUX and Digital Modulation*

Elements of the extension field GF(16) are mapped into symbols of a 16-QAM constellation. We assume a 2D-Gray mapping [deO & BAT 92], which minimise the bit error rate over additive white Gaussian noise and provide a simple detection algorithm. Points of the transmitted sequence $V_{comp} = (1, 0, \alpha^{10}, \alpha^5, \alpha^{10})$ are highlighted (red-cross points) in figure 2.

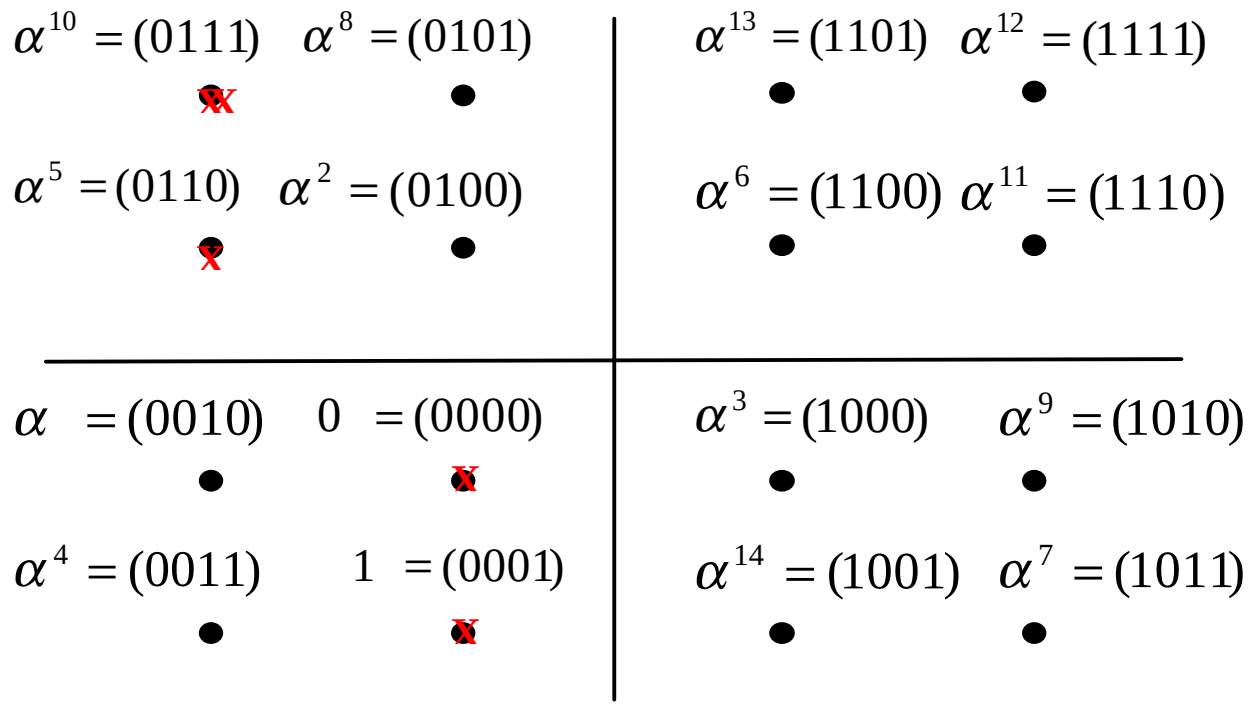

Fig. 2 – Signal constellation and GF(16) elements map.

It is worthwhile to remark that another digital modulation could be used to carry the Galois field symbols. For instance, assuming the constellation as a BPSK, an amount of 60 symbols per frame should be transmitted so as to mux 15 users. When cyclotomic compression is adopted, only 20 symbols per frame are required to transmit the information of all users. Table IV illustrates some of involved trade-offs when the constellation is changed, fixed the number of users to be N = 15 over GF(16).

Table IV – N = 15 spreading sequences using different kinds of digital modulation.

| digital modulation | symbols per frame | BW requirements |
|---|---|---|
| BPSK | 60(20) | $20B_1$ (5/3 $B_{TDM}$) |
| QPSK | 30(10) | $10B_1$ (2/3 $B_{TDM}$) |
| 16-QAM | 15(5) | $5B_1$ (1/3 $B_{TDM}$) |

### 3.3 *Retrieval of Missing Spectral Components*

The spectral decompression can be carried out by assuming that no errors occur. This can be done by using properties of valid spectra [BLA 79]. Over a field of characteristic p, the FFFT valid spectra hold $V_k^p = V_{pk} (\bmod N)$. In example 1, $V_k^2 = V_{2k} (\bmod 15)$, so that:

C(0): $V_0^2 = V_{0\,(\bmod 15)} = 1$
C(1): $V_1^2 = V_{2\,(\bmod 15)} \rightarrow V_2 = 0$; $V_2^2 = V_{4\,(\bmod 15)} = 0$
$V_4^2 = V_{8\,(\bmod 15)} \rightarrow V_8 = 0$; $V_8^2 = V_{1\,(\bmod 15)}$
C(3): $V_3^2 = V_{6\,(\bmod 15)} \rightarrow V_6 = \alpha^5$; $V_6^2 = V_{12\,(\bmod 15)} = \alpha^{10}$
$V_{12}^2 = V_{9\,(\bmod 15)} \rightarrow V_9 = \alpha^5$; $V_9^2 = V_{3\,(\bmod 15)}$
C(5): $V_5^2 = V_{10\,(\bmod 15)} \rightarrow V_{10} = \alpha^{10}$; $V_{10}^2 = V_{5\,(\bmod 15)}$
C(7): $V_7^2 = V_{14(\bmod 15)} \rightarrow V_{14} = \alpha^5$; $V_{14}^2 = V_{13(\bmod 15)} = \alpha^{10}$
$V_{13}^2 = V_{11\,(\bmod 15)} \rightarrow V_{11} = \alpha^5$; $V_{11}^2 = V_{7(\bmod 15)}$.

The (full) spectrum V is promptly recovered as:

$$V = (1, 0, 0, \alpha^{10}, 0, \alpha^5, \alpha^5, \alpha^{10}, 0, \alpha^5, \alpha^{10}, \alpha^5, \alpha^{10}, \alpha^{10}, \alpha^5).$$

### 3.4 *Despreading Galois-Carriers*

Once the spreading is carried out by the FFFT, the "despreading" corresponds precisely to the inverse-FFFT of length N | $p^m$ – 1. The $FFFT^{-1}$ allows recovering the users information, that is:

$$v_i = \frac{1}{N(\bmod p)} \sum_{k=0}^{N-1} V_k \alpha^{-ik}$$

from which the original messages may be extracted. Message data from the tributaries can be directly obtained by the scheme in figure 3. The recovered vector V' is then multiplied by a sequence of length N with all components identical to 1 / N (mod p). The sequences $\{\alpha^{-ik}\}$ have components equal to the multiplicative inverse over GF($p^m$) of the components of the Galois-carriers.

<u>Example 2</u>: (Shortened-Galois-Fourier carriers). Since N | $p^m$ –1, it would be possible to multiplex a number of N = {1, 3, 5, 15} users over GF(16). Let $\beta = \alpha^3$ be an element of order 5 over GF(16). Table V shows elements of shortened-sequences.

Table V – Galois Field GF(16).

| i | $\alpha^i$ | assoc. vector | ord ($\alpha^i$) | minimal polynomial |
|---|---|---|---|---|
| 0 | $\beta^0=(\alpha^3)^0=1$ | (0,0,0,1) | 1 | x + 1 |
| 1 | $\beta^1=(\alpha^3)^1=\alpha^3$ | (0,1,0,0) | 5 | $x^4 + x^3 + x^2 + x + 1$ |
| 2 | $\beta^2=(\alpha^3)^2=\alpha^6=\alpha^3+\alpha^2$ | (1,1,0,0) | 5 | $x^4 + x^3 + x^2 + x + 1$ |
| 3 | $\beta^3=(\alpha^3)^3=\alpha^9=\alpha^3+\alpha$ | (1,0,1,0) | 5 | $x^4 + x^3 + x^2 + x + 1$ |
| 4 | $\beta^4=(\alpha^3)^4=\alpha^{12}=\alpha^3+\alpha^2+\alpha+1$ | (1,1,1,1) | 5 | $x^4 + x^3 + x^2 + x + 1$ |

In this case, the Galois-Fourier carriers are: $\beta^{0k}=(1,1,1,1,1)$; $\beta^{1k}=(1,\alpha^3,\alpha^6,\alpha^9,\alpha^{12})$; $\beta^{2k}=(1,\alpha^6,\alpha^{12},\alpha^3,\alpha^9)$; $\beta^{3k}=(1,\alpha^9,\alpha^3,\alpha^{12},\alpha^6)$; $\beta^{4k}=(1,\alpha^{12},\alpha^9,\alpha^6,\alpha^3)$. Correlation properties (Proposition 1) remain valid since ord($\beta$) = 5. Supposing now that during an interval T, the data to be transmitted from the 5-users are:

$$v = (0, 1, 1, 0, 1)$$

We have then, V = (1, $\alpha^7$, $\alpha^{14}$, $\alpha^{11}$, $\alpha^{13}$). This time, the cyclotomic cosets are C0 = (0) and C1 = (1, 2, 4, 3). The spectral compression generates a vector $V_{comp}$ = (1, $\alpha^7$). The compression gain has a slight change. We have $\gamma_{cc}$= 2,5. Using the valid spectra properties we get the complete spectral recovering.

C(0): $V_0^2 = V_{0\,(\bmod 5)} = 1$
C(1): $V_1^2 = V_{2\,(\bmod 5)} \rightarrow V_2 = \alpha^{14}$; $V_2^2 = V_{4\,(\bmod 5)} = \alpha^{13}$
$V_4^2 = V_{3\,(\bmod 5)} \rightarrow V_3 = \alpha^{11}$; $V_3^2 = V_{1\,(\bmod 5)}$,

so that V = (1, $\alpha^7$, $\alpha^{14}$, $\alpha^{11}$, $\alpha^{13}$) as expected. The mapping and the modulation technique are the same as in example 1 despite the replacement of spreading sequences by shortened-spreading sequences.

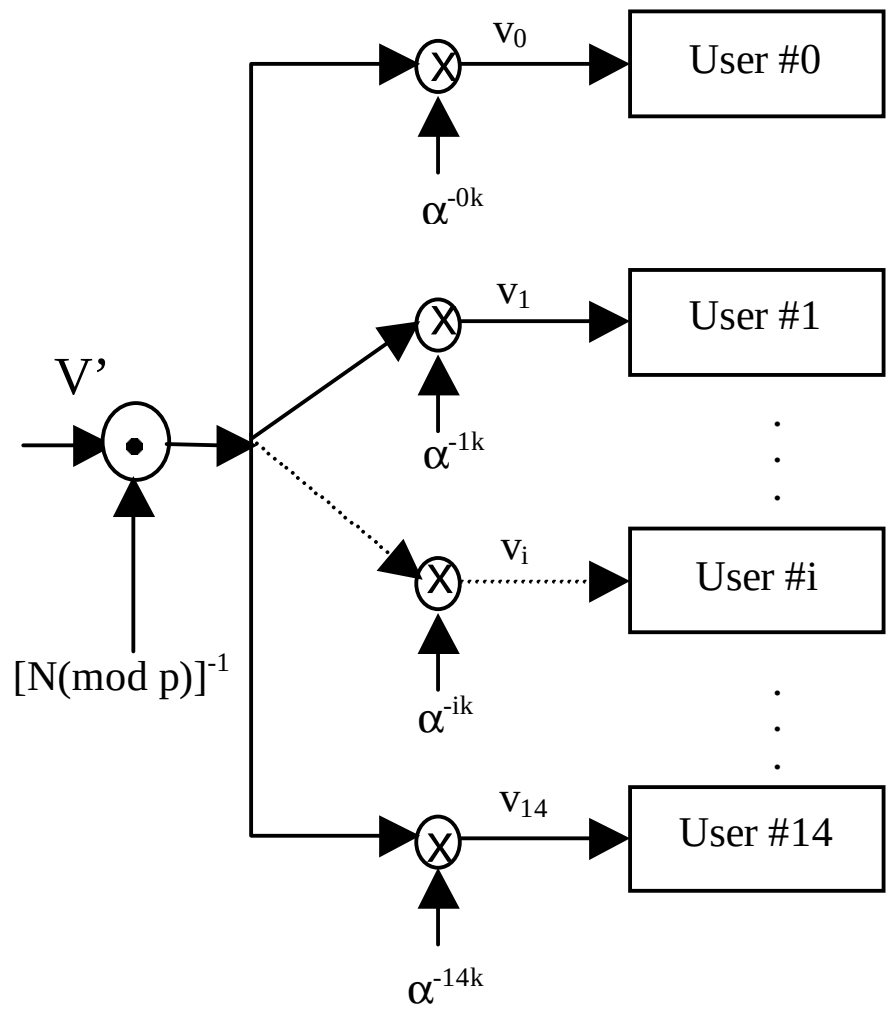

Fig. 3 – Despreading using Galois-Fourier carriers over GF(16).

## 4. GDM Systems Performance

Let $P_M$ be the probability of a bit error. Thus:

$$P_C = 1 - P_M$$

The probability of a GDM symbol error is the probability that one error occurs in a "user"-vector of length N (i.e., a GDM frame error). Then one has:

$$P_E = 1 - (1 - P_M)^N$$

Closed formulas for $P_M$ can be easily found for the most common modulation techniques [PRO 95]. The performance of N-GDM systems was analytically evaluated for N = 15 and N = 5 (shortened), both over GF(16). Effects of cyclotomic compression were investigated as a second step, for which:

$$P_E = 1 - (1 - P_M)^{\nu}$$

Results are plotted in figures 4 – 7 below.

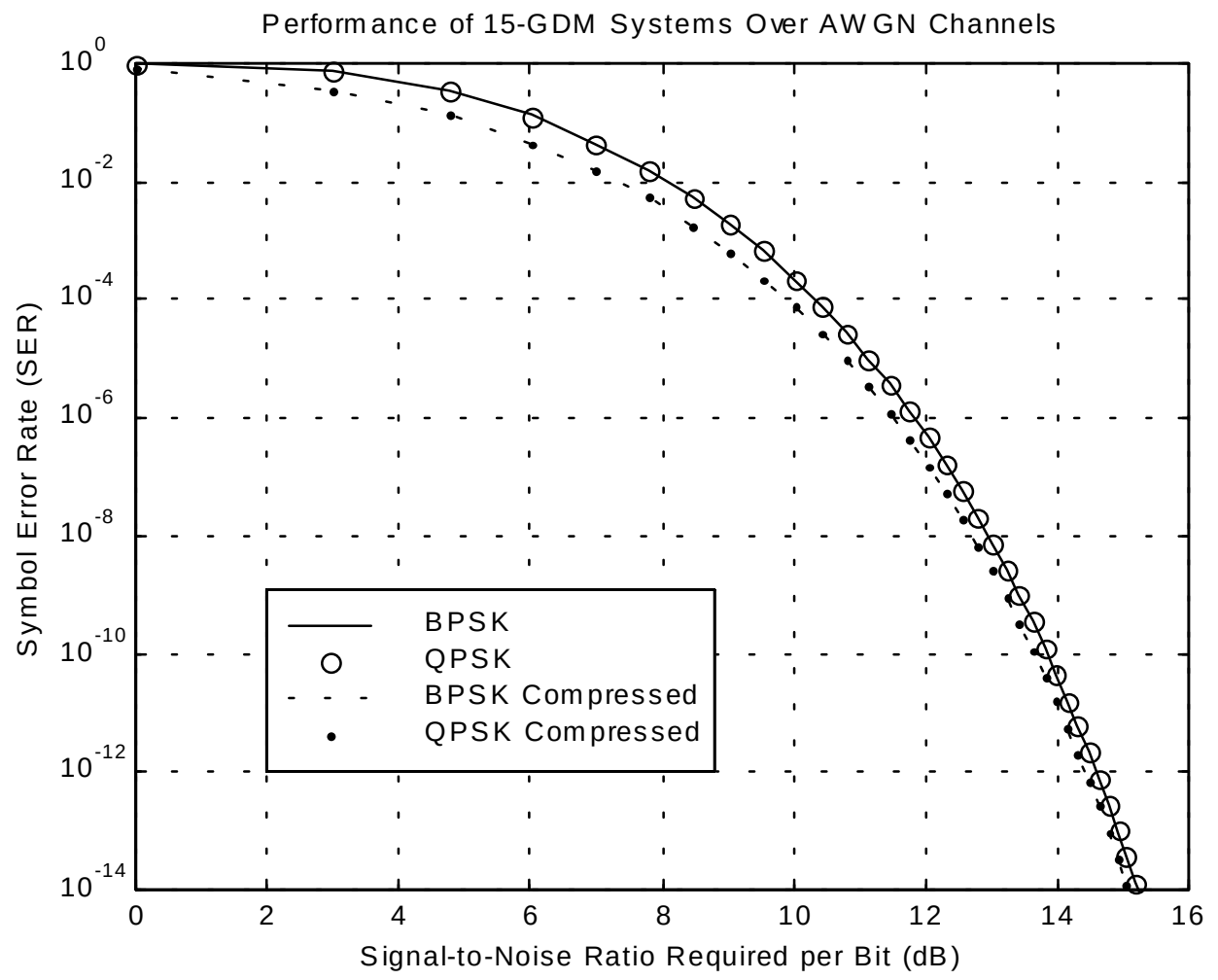

Fig. 4 – Analytical SER estimation for 15-user GDM based on the FFFT over AWGN channels (BPSK and QPSK modulation.)

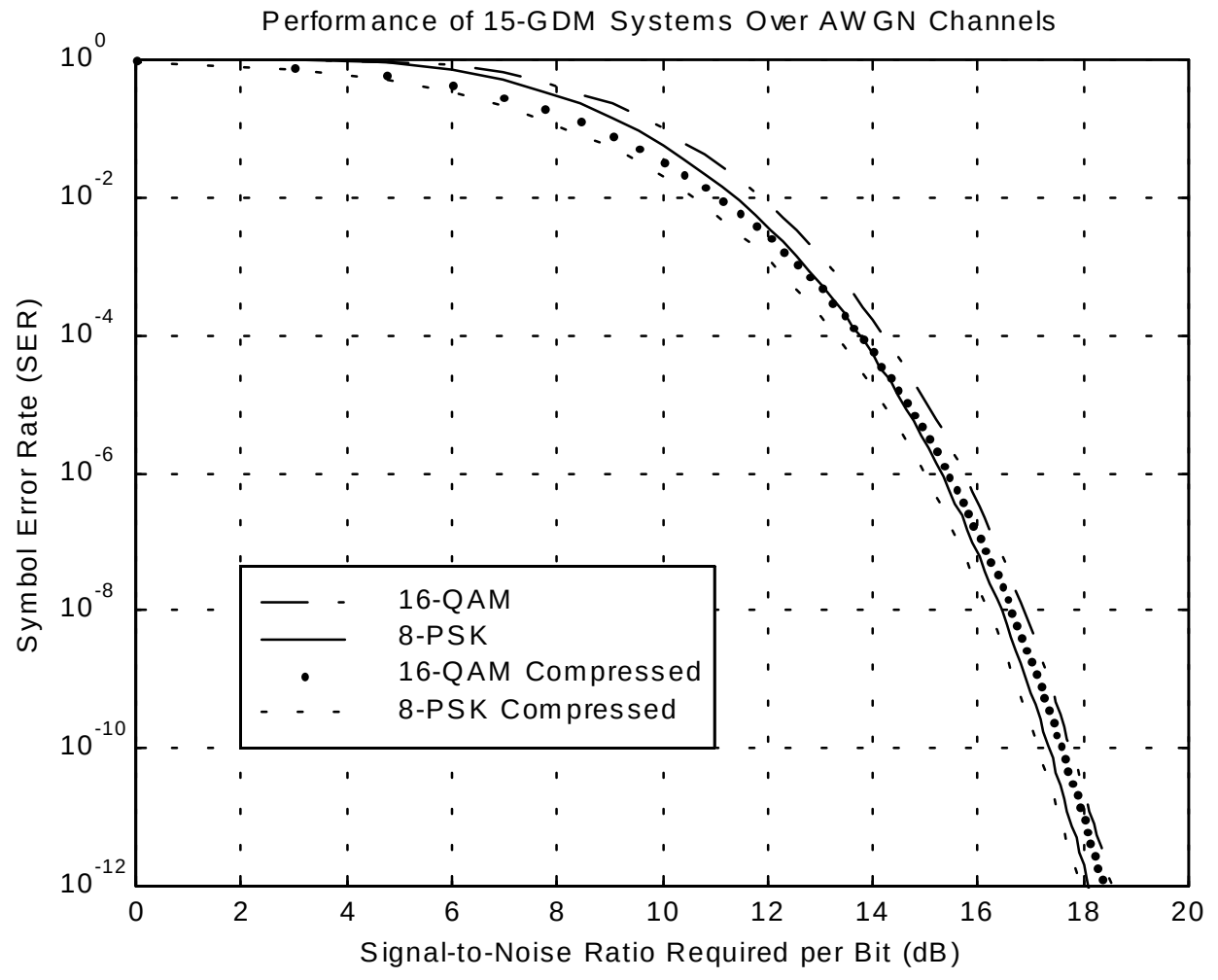

Fig. 5 – Analytical SER estimation for 15-user GDM based on the FFFT over AWGN channels. (8-PSK and 16-QAM modulation.)

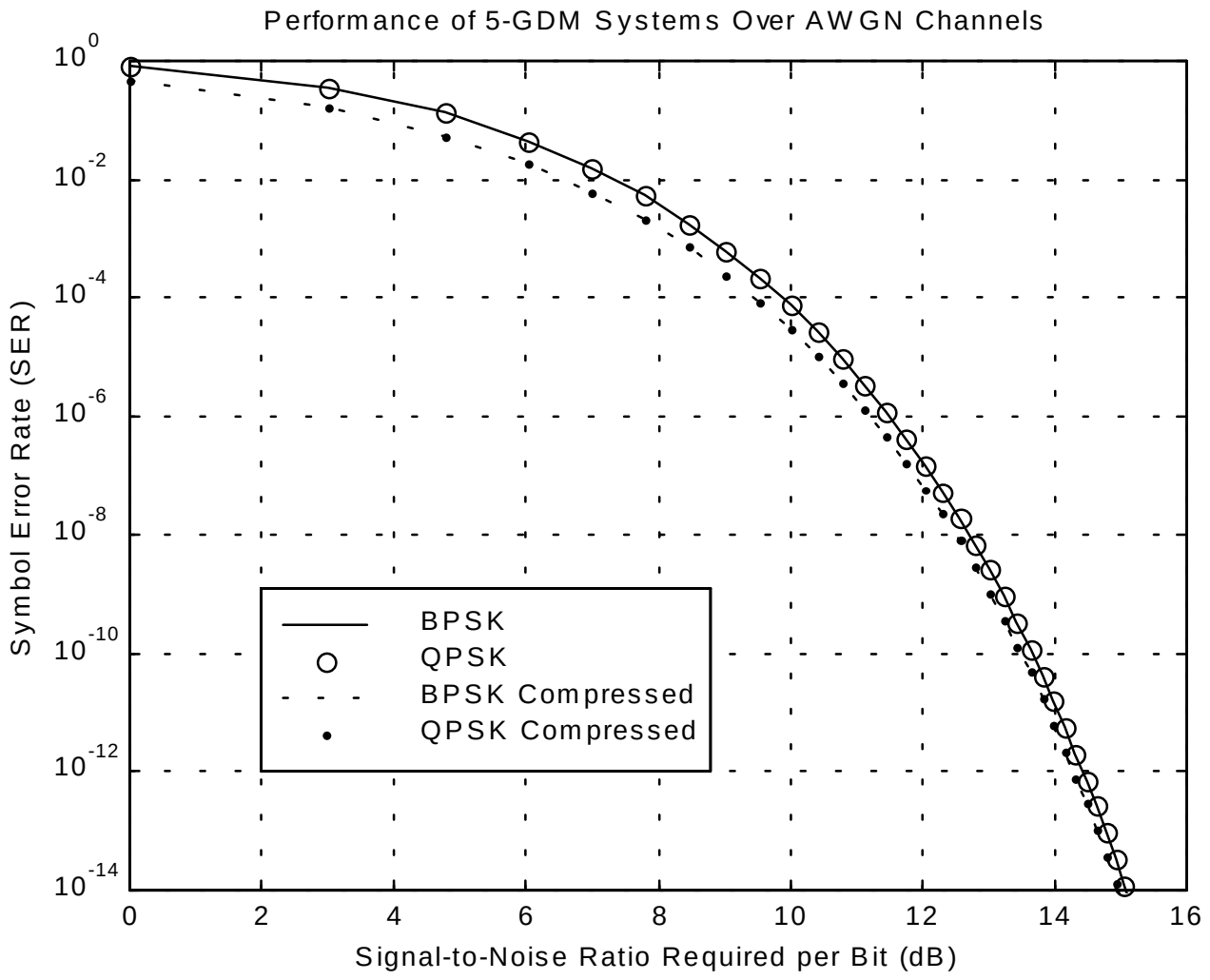

Fig. 6 – Analytical SER estimation for 5-user GDM based on the FFFT over AWGN channels. (BPSK and QPSK modulation.)

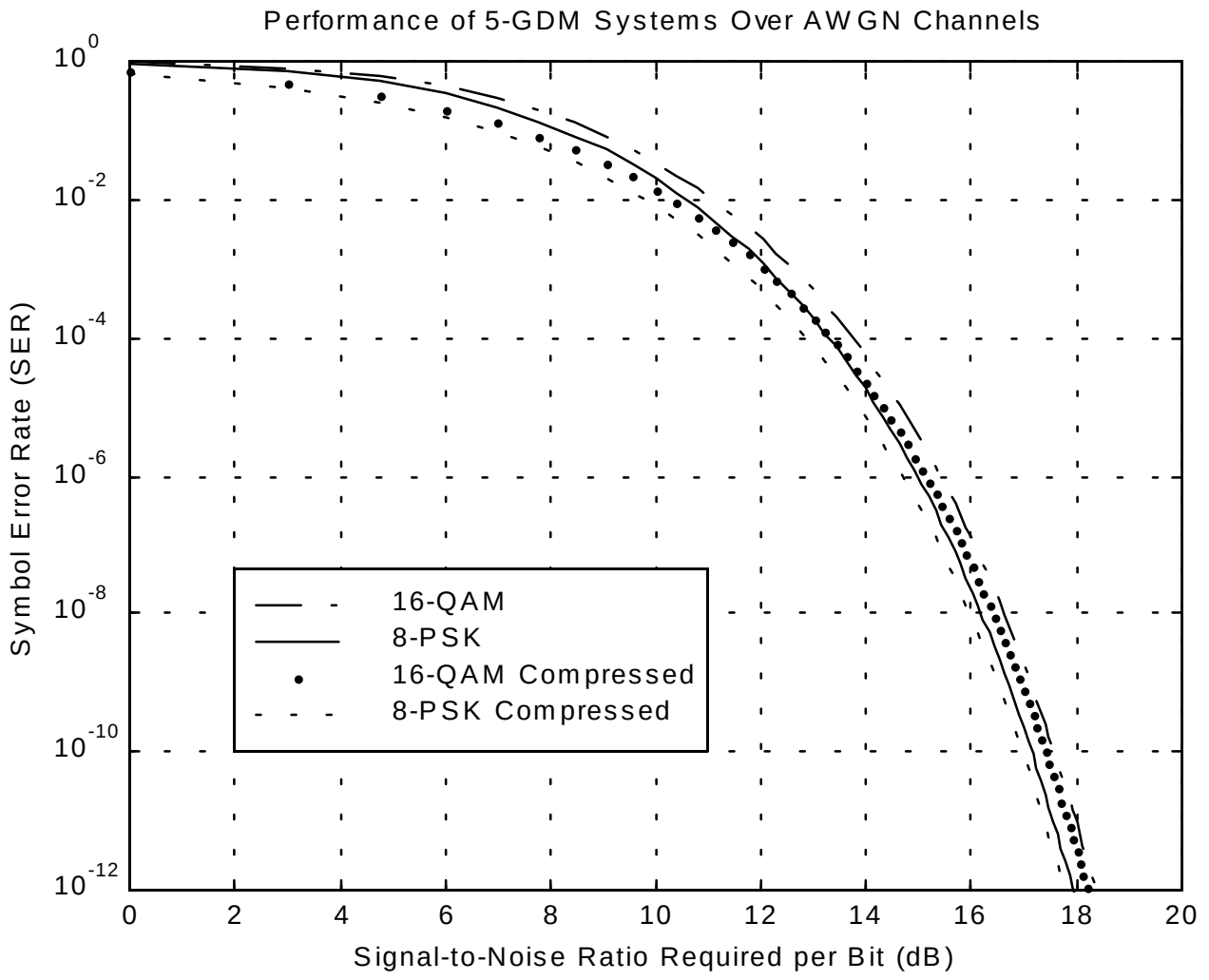

Fig. 7 – Analytical SER estimation for 5-user GDM based on FFFT over AWGN channels. 8-PSK and 16-QAM modulation applied.

Usually, QPSK is the most common technique in commercial CDMA systems. However, the application of other modulation techniques such as 16-QAM or 8-PSK in high data rate transmissions over CDMA networks were investigated in a recent paper [BEN et al. 99].

## 5. Concluding Remarks

Finite Fields Transforms are offered as a new tool of spreading sequence design. New digital multiplex schemes based on such transforms have been introduced which are multilevel Code-Division Multiplex. They are attractive due to their better spectral efficiency regarding classical FDM / TDM which require a bandwidth expansion roughly proportional to the number of channels to be multiplexed. This new approach is promising for communication channels supporting a high signal-to-noise ratio. A number of practical matters such as imperfect synchronisation, bit error rate (BER) estimation, or unequal user power are left to be investigated. Tributaries are rather stacked instead of time or frequency interleaved, yielding new "Efficient-Spread-Spectrum Sequences for Band-Limited Channels".

## Acknowledgements

The authors thank Professor Paddy Farrell for his constructive criticism.